\documentclass[superscriptaddress]{revtex4} 
 \usepackage{graphicx}
 \usepackage{color}
 \usepackage[sort&compress]{natbib}
 \usepackage{mathrsfs} 
 \usepackage{lscape,epsfig}
 \usepackage{amsmath}
 \usepackage{bm}
 \usepackage{txfonts}
 \usepackage{multirow}

  \def\beq{\begin{equation}}
  \def\eeq{\end{equation}}
  \def\beqa{\begin{eqnarray}}
  \def\eeqa{\end{eqnarray}}
  \def\ban{\begin{eqnarray*}}
  \def\ean{\end{eqnarray*}}
  \def\bi{\begin{itemize}}
  \def\ei{\end{itemize}} 

\begin{document}

\title{Nuclear matter properties from local chiral interactions with $\Delta$ isobar intermediate states} 
\author{Domenico Logoteta}
\affiliation{INFN, Sezione di Pisa, Largo Bruno Pontecorvo 3, I-56127 Pisa, Italy} 
\author{Ignazio Bombaci}
\affiliation{Dipartimento di Fisica, Universit\'a di Pisa, Largo Bruno Pontecorvo 3, I-56127 Pisa, Italy }
\affiliation{INFN, Sezione di Pisa, Largo Bruno Pontecorvo 3, I-56127 Pisa, Italy}
\affiliation{European Gravitational Observatory, Via E. Amaldi, I-56021 S. Stefano a Macerata, Cascina Italy}
\author{Alejandro Kievsky}
\affiliation{INFN, Sezione di Pisa, Largo Bruno Pontecorvo 3, I-56127 Pisa, Italy}

%
\begin{abstract}
Using two-nucleon and three-nucleon interactions derived in the framework of chiral perturbation theory (ChPT) 
with and without the explicit $\Delta$ isobar contributions, we calculate the energy per particle of 
symmetric nuclear matter and pure neutron matter in the framework of the microscopic Brueckner--Hartree--Fock approach. 
In particular, we present for the first time nuclear matter calculations using the new fully local in 
coordinate-space two-nucleon interaction at the next-to-next-to-next-to-leading-order (N3LO) of ChPT 
with $\Delta$ isobar intermediate states (N3LO$\Delta$) recently developed by Piarulli {\it et al.} [arXiv:1606:06335]. 
We find that using this N3LO$\Delta$ potential, supplemented with a local N2LO three-nucleon interaction with 
explicit $\Delta$ isobar degrees of freedom, it is possible to obtain a satisfactory saturation point of 
symmetric nuclear matter.
For this combination of two- and three-nucleon interactions we also calculate  
the nuclear symmetry energy and we compare our results with the empirical constraints on this quantity 
obtained using the excitation energies to isobaric analog states in nuclei and using experimental data on the 
neutron skin thickness of heavy nuclei, finding a very good agreement with these empirical constraints in all 
the considered nucleonic density range. 
In addition, we find that the explicit inclusion of $\Delta$ isobars diminishes the strength of the 
three-nucleon interactions needed the get a good saturation point of symmetric nuclear matter.    
We also compare the results of our calculations with those obtained by other research groups using 
chiral nuclear interactions with different many-body methods, finding in many cases a very satisfactory agreement.    
\end{abstract}

\maketitle
  
\vspace{0.50cm}
  PACS number(s): {21.65.-f,   
                   21.30.-x,   
                   21.65.Ef,   
                   26.60.Kp    
                           } 
\vspace{0.50cm}
  
\section{Introduction}

Effective field theory (EFT) for low-energy quantum chromodynamics (QCD) opened a new and systematic way 
to describe the nuclear interaction \cite{weimberg} (for comprehensive and thorough reviews on this subject 
see Refs.~\cite{epel06,epel09,machl11,holt13}). 
The considerable advantage of using such method lies in the fact that two-body, three-body as well as 
many-body nuclear interactions can be calculated perturbatively, i.e. order by order, according to 
a well defined scheme based on a low-energy effective QCD Lagrangian which retains the symmetries of QCD, 
and in particular the approximate chiral symmetry. 
This chiral perturbation theory (ChPT) is based on a power counting in the ratio $Q/\Lambda_\chi$, 
where $Q$ denotes a low-energy scale identified with the magnitude of the three-momenta of the external nucleons and with the pion mass $m_\pi$ whereas $\Lambda_\chi \sim 1~{\rm GeV}$ denotes 
the chiral symmetry breaking scale. 
In this framework single-pion exchange and multi-pion exchange give the long- and intermediate-range part 
of the nuclear interaction, whereas the short-range component is included via nucleon contact terms.   
Within this approach the details of the QCD dynamics are contained in parameters, the so called low-energy 
constants (LECs), which are fixed by low-energy experimental data.
This systematic technique is especially fruitful in the case of nucleonic systems where the relevance 
of the three-nucleon force (TNF) is a well established feature \cite{kalantar12,hammer13,binder16}.    

Nucleon-nucleon (NN) interactions plus TNFs based on ChPT have been recently used to investigate 
properties of medium-mass nuclei \cite{ekstrom,roth} and heavy nuclei \cite{binder}.  
A very important task in this line, is the evaluation of the uncertainties originating in the nuclear 
Hamiltonian \cite{furnstahl15} and in particular on the LECs and to establish which should be the 
best fitting procedure to fix them \cite{furn15}.  
For example, in Ref.\ \cite{ekstrom} a simultaneous optimization of the NN interaction plus a TNF
in light and medium-mass nuclei has been performed.  

The standard version of ChPT is based on pion and nucleon degrees of freedom only. 
However, due to its low excitation energy $\Delta m = m_\Delta - m_N = 293.1~{\rm MeV}$ and to the strong coupling with the pion-nucleon ($\pi$N) system, as it is well known, the $\Delta(1232)$ isobar (hereafter the $\Delta$ isobar) plays 
an important role in nuclear physics.   
In the standard, i.e. $\Delta$-less, ChPT the effects of the $\Delta$ isobar are taken into account 
implicitly and are incorporated in LECs (such as $c_3$ and $c_4$) which are fixed using $\pi$N scattering data or by fitting NN phase shifts. 
Alternatively, since $\Delta m << \Lambda_\chi$, one could extend ChPT to include the $\Delta$ isobar  
as an additional and explicit degree of freedom. 
It has been shown by various authors \cite{kaiser98,krebs07} 
that this $\Delta$-full ChPT has an improved convergence with respect to the  $\Delta$-less ChPT. 
In addition, the $\Delta$-full ChPT naturally leads to TNFs induced by two-pion exchange with excitation 
of an intermediate $\Delta$ (the celebrated Fujita--Miyazawa three-nucleon force \cite{fujita57}). 

Recently Piarulli et al. \cite{maria_local} have developed a fully local in coordinate-space two-nucleon  
chiral potential which includes the $\Delta$ isobar intermediate state. This new potential represents 
the fully local version of the minimally non-local chiral interaction reported in Ref.~\cite{maria15}. 
Local coordinate space interactions are particularly suitable for Quantum Monte Carlo calculations of 
nuclei and nuclear matter. 

In this work, we present for the first time microscopic calculations of the equation of state (EOS) 
of symmetric nuclear matter and pure neutron matter using the local chiral potential of Ref.\ \cite{maria_local} 
and employing the Brueckner--Bethe--Goldstone (BBG) \cite{bbg1,bbg2} many body theory within the 
Brueckner--Hartree--Fock (BHF) approximation.  
The present work represents a development with respect to our previous works \cite{logoteta15,logoteta16} 
where ChPT nuclear interactions have been used in BHF calculations of nuclear matter properties.  
Nuclear interactions based on ChPT have been also used by several groups for calculating the EOS 
of pure neutron matter \cite{hebeler10,darm13,rogg14,wlaz14,gand15,AFDMC,SGR,drischler16} and 
symmetric nuclear matter \cite{hebeler11,li-schu_12,arianna13,coraggio14,kohno15}. 
A comparison with some of these calculations will be performed in last part of this work. 

The paper is organized as follows: 
in section 2 we review the two-body and three-body chiral interactions used in our calculations; 
in section 3  we briefly recall the basic features of the BBG many-body theory and we discuss 
how to include a TNF in the BHF approach;  
section 4 is devoted to show and discuss the results of our calculations; 
finally in section 5 we summarize our results and outline the main conclusions of the present study.

\section{Chiral nuclear interactions}

We consider interactions fully derived in ChPT both for the two- and three-nucleons sectors.  
We use indeed NN potentials calculated at the next-to-next-to-next-to-leading-order (N3LO) of ChPT, 
in conjunction with three-nucleon interactions at the next-to-next-to-leading-order (N2LO). 
Currently, chiral NN potentials have been calculated up to order N4LO by Epelbaum et al.~\cite{epeln4lo} and 
by Entem et al. \cite{entemn4lo}. 
In addition, dominant contribution at order N5LO have been analyzed in Ref.\ \cite{entemn5lo} where it was found 
a satisfactory convergence of the perturbative expansion of the NN potential.  
Also chiral TNFs have been calculated at higher orders of the perturbative expansion. 
In Ref. \cite{N3LO_3} the N3LO contributions to the TNFs have been derived.  
Sub-leading contributions to the TNFs appear at order N4LO \cite{girlandan4lo} and they may be 
potentially important. 
The use of these large variety of forces in practical calculations is not an easy task and many efforts 
are currently devoted to incorporate part of these interactions in studies of few- and many-body systems.  
In this work we restrict our calculations to the use of chiral TNFs evaluated at order N2LO.   

Let us now focus on the specific interactions we have employed in the present work. 
As a two-body nuclear interaction, we have used the fully local chiral potential at N3LO including $\Delta$ isobar excitations in intermediate state (hereafter N3LO$\Delta$) recently proposed in Ref.\ \cite{maria_local}.  
Originally this potential was presented in Ref.\ \cite{maria15} in a minimal non-local form. 
The first chiral NN potential in local form has been derived in Ref. \cite{gezerlis_loc} 
by Gezerlis et al. considering a $\Delta$-less ChPT up to order N2LO. 
Notice that Ref.\ \cite{maria_local} reports different parametrizations of the local potential obtained 
fitting the low energy NN experimental data using different long- and short-range cutoffs.   
In the calculations presented in this work, we use the {\it model~b} described in Ref.\ \cite{maria_local} 
(see their Tab. II) which fits the Granada database \cite{granada} of proton-proton ($pp$) and 
neutron-proton ($np$) scattering data up to an energy of $125$ MeV in the laboratory reference frame and has 
a $\chi^2$/datum$\sim$ $1.07$.

There are several indications that $\Delta$ isobar plays an important role in nuclear processes. 
For instance the excitation of $\Delta$ isobar is needed to reproduce the observed energy spectra of 
low-lying states in $s$- and $p$-shell nuclei as well as the correct spin-orbit splitting in low-energy 
$n$-$\alpha$ scattering. 
It is therefore very important to use this new chiral potential \cite{maria_local} also in nuclear matter calculations. 

We have also used the N3LO chiral NN potential by Entem and Machleidt (EM) \cite{N3LO}, considering two 
different values of the cutoff, $\Lambda = 500$~MeV and $\Lambda = 450$~MeV, 
employed to regularize the high momentum components of the interaction. 
Notice that for consistency reasons, the same value of the cutoff has been employed in each calculation,   
both in the two- and three-nucleon interactions. However the assumed shape of the cutoff in the two-body 
and in the three-body interaction is in general different (see next section for more details).

Concerning the TNF, we have used the N2LO potential by Epelbaum et al.~\cite{N2LO}) in its local version 
given by Navratil \cite{N2LOL}. We note that the non locality of the N2LO three-nucleon interaction 
depends only on the cutoff used to regularize the potential. 
The N2LO TNF force has the following structure in momentum space: 
\begin{equation}
V_{NNN}^{(2\pi)} = \sum_{i\neq j\neq k} \frac{g_A^2}{8f_\pi^4} 
\frac{\bm{\sigma}_i \cdot \bm{q}_i \, \bm{\sigma}_j \cdot
\bm{q}_j}{(\bm{q_i}^2 + m_\pi^2)(\bm{q_j}^2+m_\pi^2)}
F_{ijk}^{\alpha \beta}\tau_i^\alpha \tau_j^\beta,
\label{nnn1}
\end{equation}
\begin{equation}
V_{NNN}^{(1\pi)} = -\sum_{i\neq j\neq k}
\frac{g_A c_D}{8f_\pi^4 \Lambda_\chi} \frac{\bm{\sigma}_j \cdot \bm{q}_j}{\bm{q_j}^2+m_\pi^2}
\bm{\sigma}_i \cdot
\bm{q}_j \, {\bm \tau}_i \cdot {\bm \tau}_j ,
\label{nnn2}
\end{equation}
\begin{equation}
V_{NNN}^{(\rm ct)} = \sum_{i\neq j\neq k} \frac{c_E}{2f_\pi^4 \Lambda_\chi}
{\bm \tau}_i \cdot {\bm \tau}_j,
\label{nnn3}
\end{equation}
where $\bm{q}_i=\bm{p_i}^\prime -\bm{p}_i$ is the difference between the 
final and initial momentum of nucleon $i$ and 
\begin{equation}
F_{ijk}^{\alpha \beta} = \delta^{\alpha \beta}\left (-4c_1m_\pi^2
 + 2c_3 \bm{q}_i \cdot \bm{q}_j \right ) + 
c_4 \epsilon^{\alpha \beta \gamma} \tau_k^\gamma \bm{\sigma}_k
\cdot \left ( \bm{q}_i \times \bm{q}_j \right ).
\label{nnn4}
\end{equation}
In equations (\ref{nnn1})--(\ref{nnn4}) $\bm\sigma_i$ and $\bm\tau_i$ are Pauli matrices for spin and 
isospin spaces while $g_A=1.29$ and $f_\pi = 92.4$ MeV are the axial-vector coupling and the pion decay constant. 
The nucleon labels $i$, $j$, $k$ can take values 1, 2, 3, which results in six possible permutations in each sum.   
Factors $c_1$, $c_3$, $c_4$, $c_D$ and $c_E$ are the low energy constants.  
The interaction described above keeps the same operatorial structure both including or not the $\Delta$ 
degrees of freedom (\cite{krebs07}).   
We note that the constants $c_1$, $c_3$ and $c_4$ entering in Eq.\ (\ref{nnn4})
 are already fixed at two-body level 
by the N3LO interaction.    
However when including the $\Delta$ isobar in the three-body potential, the parameters $c_3$ and $c_4$ 
take additional contribution from the Fujita--Miyazawa diagram. 
Such a diagram appears at order NLO and is clearly not present in the theory without the $\Delta$. 
In order to properly take into account this contribution, one has to add to the values of $c_3$ and $c_4$ (\cite{maria15}), the quantity given by the relation 
\cite{krebs07}: $c_3^{\Delta}=-2 c_4^{\Delta}=-\frac{h_A^2}{9 \Delta m}$ being $h_A=\frac{3 g_A}{\sqrt{2}}=2.74$ 
and $\Delta m = m_\Delta - m_N = 293.1$ MeV, where $m_\Delta$, $m_N$ are the $\Delta$ isobar and the nucleon masses and $h_A$ is the leading order $\Delta N \pi$ axial-vector coupling constant.  
Because of this, we have for the TNF: $c_3=\tilde c_3+c_3^{\Delta}$ and $c_4=\tilde c_4+c_4^{\Delta}$ being  
$\tilde c_3$ and $\tilde c_4$ the values fixed at two-body level. For the N3LO$\Delta$ potential we have: $\tilde c_3=0.79$ GeV$^{-1}$ and $\tilde c_4=1.33$ GeV$^{-1}$ \cite{maria15}. 
For the EM potential \cite{N3LO}, $c_i$ and $\tilde c_i$ coincide because there is no additional diagram to sum up.   
The values of the constants $c_i$ for the TNFs that we have considered in the present work are reported in 
Tab. \ref{tab1}. 

The remaining parameters $c_D$ and $c_E$ are not determined by the two-body interaction and have to 
be fixed constraining some specific observable of few-body nuclear systems or to reproduce the 
empirical saturation point of symmetric nuclear matter.   
In particular, for the interaction model N3LO+N2LO(450), following reference \cite{marcucci12},  
we have set $c_D=-0.24$ and $c_E=-0.11$; these values are able to reproduce binding energies of $^3$H and 
$^3$He and the Gammow-Teller matrix element for the $^3$H $\beta$-decay considering contributions to the 
axial nuclear current up to order N3LO \cite{marcucci12}.   
For the interaction model N3LO+N2LO(500), we have adopted a recent constraint on $c_D$ and $c_E$ employing the same 
strategy of Ref.\ \cite{marcucci12} but considering contributions to the axial nuclear current up to 
order N4LO \cite{baroni16}. 
We note that this parametrization has also the valuable property to reproduce 
the neutron--deuteron doublet scattering length.  

Finally, for very recent model N3LO$\Delta$+N2LO$\Delta$ \cite{maria_local} no calculation for few-body 
nuclear systems has been done so far.  
Thus we have fitted the LECs $c_D$ and $c_E$ to get a good saturation point for symmetric nuclear matter. 
We want to point out that different values of $c_D$ and $c_E$ can produce equally satisfactory nuclear 
matter saturation points. In order to explore this possibility, we provide two parametrizations of the 
N2LO$\Delta$ TNF able to fulfill this constraint (see Tab.\ \ref{tab2}).   
Hereafter we refer to these two parametrizations of the three-body interaction as 
the N2LO$\Delta1$ and N2LO$\Delta2$ models.      
However, it is important to note that depending on the particular choice adopted to fit the TNF, 
the properties of the energy per particle at large density may change considerably. 
We shall discuss this point in next sections. 

 \begin{table} 
\begin{tabular}{l|cccccc}
\hline
      TNF            & $c_D$ &   $c_E$  & $c_1$    & $c_3$   & $c_4$   \\               
\hline
  N2LO$\Delta1$      & -0.10 &   1.30   & -0.057 & -3.63 & 3.14  \\ 
  N2LO$\Delta2$      & -4.06 &   0.37   & -0.057 & -3.63 & 3.14  \\ 
  N2LO500            & -1.88 &  -0.48   & -0.810 & -3.20 & 5.40   \\      
  N2LO450            & -0.11 &  -0.24   & -0.810 & -3.40 & 3.40   \\
\hline
\hline
 \end{tabular}
\caption{Values of the low energy constants (LECs) of the TNFs models used in the present calculations. 
In the first and in the second row, we report the parametrizations of the N2LO three-body force 
with the $\Delta$ isobar excitations \cite{maria_local}. Notice that the values $c_1$, $c_3$ and $c_4$ 
have been kept fixed. 
In the third and in the forth rows we report the N2LO TNF parametrizations obtained in conjunction with 
the EM \cite{N3LO} N3LO two-nucleon potential with $\Lambda=500$ MeV (third row) and with 
$\Lambda=450$ MeV (forth row). 
The LECs $c_1$, $c_3$ and $c_4$ are expressed in GeV$^{-1}$, whereas $c_D$ and $c_E$ are dimensionless. 
} 
\label{tab1}
\end{table} 

\section{The BHF approach with averaged three-body forces} 

The Brueckner--Hartree--Fock (BHF) approach is the lowest order of the Brueckner--Bethe--Goldstone (BBG) 
many-body theory \cite{bbg1,bbg2}.   
In this theory, the ground state energy of nuclear matter is evaluated in terms of the so-called hole-line 
expansion, where the perturbative diagrams are grouped according to the number of independent hole-lines. 
The expansion is derived by means of the in-medium two-body scattering Brueckner $G$-matrix  
which describes the effective interaction between two nucleons in presence of the surrounding nuclear medium.  
In the case of asymmetric nuclear matter 
{\footnote{In the present work we consider spin unpolarized nuclear matter. Spin polarized nuclear matter 
within the BHF approach has been considered, for example, in Ref. \cite{vb02,bomb+06}.}
with neutron density $\rho_n$, proton density $\rho_p$, total nucleon density $\rho = \rho_n + \rho_p$ 
and isospin asymmetry $\beta= (\rho_n-\rho_p)/\rho$ (asymmetry parameter), 
one has different G-matrices describing the $nn$, $pp$ and $np$ in medium effective interactions.   
They are obtained by solving the well known Bethe--Goldstone equation, written schematically as 
\begin{equation}
G_{\tau_1\tau_2;\tau_3\tau_4}(\omega) = V_{\tau_1\tau_2;\tau_3\tau_4} 
 + \sum_{ij} \frac{V_{\tau_1\tau_2;\tau_i\tau_j} \ Q_{\tau_i\tau_j}}{\omega-\epsilon_{\tau_i}-\epsilon_{\tau_j} + i\varepsilon}
G_{\tau_i\tau_j;\tau_3\tau_4}(\omega) \;,
\label{bg}
\end{equation}
where $\tau_q$  ($q = 1, 2, i,j, 3, 4$) indicates the isospin projection of the two nucleons in the initial, intermediate and final states, 
$V$ denotes the bare NN interaction, $Q_{\tau_i\tau_j}$ is the Pauli operator that prevents the intermediate state nucleons $(i, j)$ from being scattered 
to states below their respective Fermi momenta $k_{F_{\tau}}$ and 
$\omega$, the so-called starting energy, corresponds to the sum of non-relativistic energies of the interacting nucleons. 
The single-particle energy $\epsilon_\tau$ of a nucleon with momentum $k$ and mass $m_\tau$ is given by
\begin{equation}
       \epsilon_{\tau}(k) = \frac{\hbar^2k^2}{2m_{\tau}} + U_{\tau}(k) \ ,
\label{spe}
\end{equation}
where the single-particle potential $U_{\tau}(k)$ represents the mean field felt by a nucleon due to its 
interaction with the other nucleons of the medium. 
In the BHF approximation, $U_{\tau}(k)$ is calculated through the real part of the so-called 
on-energy-shell $G$-matrix, and is given by
\begin{equation}
U_{\tau}(k) = \sum_{\tau'} \sum_{k'< k_{F_{\tau'}}} \mbox{Re} \ \langle k k'
\mid G_{\tau\tau';\tau\tau'}(\omega=\epsilon_{\tau}(k)+\epsilon_{\tau'}(k')) \mid k k'\rangle_A 
 \;,
\label{spp}
\end{equation}
where the sum runs over all neutron and proton occupied states and the matrix elements are properly antisymmetrized. 
We make use of the so-called continuous choice \cite{jeuk+67,gra87,baldo+90,baldo+91} for the single-particle
potential $U_{\tau}(k)$ when solving the Bethe--Goldstone equation. As shown in Refs. \cite{song98,baldo00}, 
the contribution of the three-hole-line diagrams to the energy per particle $E/A$ is minimized in this prescription and a faster convergence of the hole-line expansion for $E/A$ is achieved \cite{song98,baldo00,baldo90} with respect to the so-called gap choice for $U_{\tau}(k)$. 

Once a self-consistent solution of Eqs.\ (\ref{bg})--(\ref{spp}) is achieved, the energy per particle can be calculated as 
\begin{equation}
\frac{E}{A}(\rho,\beta)=\frac{1}{A}\sum_{\tau}\sum_{k < k_{F_{\tau}} }
 \left(\frac{\hbar^2k^2}{2m_{\tau}}+\frac{1}{2} U_{\tau}(k) \right) \ .
\label{bea}
\end{equation}

\subsection{Inclusion of three-nucleon forces in the BHF approach}
As it is well known, within the most advanced non-relativistic quantum many-body approaches,  
it is not possible to reproduce the empirical saturation point of symmetric nuclear matter,  
$\rho_{0} = 0.16 \pm  0.01~{\rm fm}^{-3}$, $E/A|_{\rho_0} = -16.0 \pm 1.0~{\rm MeV}$, 
when using two-body nuclear interactions only. 
In fact, the saturation points obtained using different NN potentials lie within a narrow 
band \cite{coester70,day81}, the so-called Coester band, with either a too large saturation density or 
a too small binding energy ($B = -E/A$) compared to the empirical values.  
In particular, SMN results over-bound with a too large saturation density when using modern high precision 
nucleon-nucleon (NN) potentials, fitting NN scattering data up to energy of $350$ MeV, with a $\chi^2$ per datum 
next to $1$ \cite{ZHLi06}. 
As in the case of few-nucleon systems \cite{kalantar12,hammer13,binder16}, also for the nuclear matter case 
TNFs are considered as the missing physical effect of the whole picture. 
The inclusion of TNF is thus required in order to reproduce a realistic saturation 
point (\cite{FP81,bbb97,apr98,Li2008,taranto13,zuo14}). 
In addition, TNFs are likely crucial in the case of dense $\beta$-stable nuclear matter to obtain a stiff 
equation of state (EOS) \cite{bbb97,apr98,li-hjs08,chamel11} compatible with the measured masses,  
$M = 1.97 \pm 0.04 \, M_\odot$ \cite{demo10} and $M = 2.01 \pm 0.04 \, M_\odot$ \cite{anto13} 
of the neutron stars in PSR~J1614-2230 and PSR~J0348+0432 respectively. 
 
Within the BHF approach TNFs cannot be used directly in their original form. 
This is because it would be necessary to solve three-body Faddeev equations in the nuclear medium 
(Bethe--Faddeev equations) \cite{bethe65,rajaraman-bethe67} and currently this is a task still far to be achieved.  
To circumvent this problem an effective density dependent two-body force is built starting from the 
original three-body one by averaging over 
one of the three nucleons \cite{loiseau,grange89}.  

In the present work, we consider the in medium effective NN force derived in Ref.\ \cite{holt}. 
The momentum space average proposed in \cite{holt} produces an effective density dependent NN potential 
of the following form:
\begin{eqnarray}
&&V_{eff}(\bm p,\bm q) = V_C + \bm \tau_1 \cdot \bm \tau_2\, W_C + 
\left [V_S + \bm \tau_1 \cdot \bm \tau_2 \, W_S \right ] \bm \sigma_1 \cdot 
\bm \sigma_2  \nonumber \\
&+& \left [ V_T + \bm \tau_1 \cdot \bm \tau_2 \, W_T \right ] 
\bm \sigma_1 \cdot \bm q \, \bm \sigma_2 \cdot \bm q \nonumber \\
&+& \left [ V_{SO} + \bm \tau_1 \cdot \bm \tau_2 \, W_{SO} \right ] \,
i (\bm \sigma_1 + \bm \sigma_2 ) \cdot (\bm q \times \bm p) \nonumber \\
&+& \left [ V_{Q} + \bm \tau_1 \cdot \bm \tau_2 \, W_{Q} \right ] \,
\bm \sigma_1 \cdot (\bm q \times \bm p)\, \bm \sigma_2 \cdot (\bm q \times \bm p)\, ,
\end{eqnarray}
where the subscripts on the functions  $V_i$, $W_i$ stand for central (C), spin (S), tensor (T), spin-orbit (SO) and quadratic spin-orbit (Q). 
Explicit expressions for these functions can be found in Ref.\ \cite{holt}. Such effective interaction was obtained closing one of the three fermion lines in the Feynman diagrams concerning the original TNF, and evaluating the resulting two-body diagram which takes into 
account the in medium modification of the nucleon propagator due to the bubble insertion. 
We also note that the same effective interaction can be obtained by averaging the original three-nucleon interaction 
$V_{NNN}$ over the generalized coordinates of the third nucleon \cite{carbone13}: 
\begin{equation}\label{eq:normord_singpart}
V_{eff} =
\text{Tr}_{(\sigma_3,\tau_3)} \int \frac{d \bm{p}_3}{(2 \pi)^3}
\, n_{\bm{p}_3} \, V_{NNN} \, (1-P_{13}-P_{23})\, ,
\end{equation}
where 
\begin{equation}
  P_{ij} = \frac{1+\bm\sigma_i\cdot\bm\sigma_j}{2} \, \frac{1+\bm\tau_i\cdot\bm\tau_j}{2} \,
         P_{\bm p_i \leftrightarrow\bm p_j}
\end{equation}
are spin-isospin-momentum exchange operators and $n_{\bm{p}_3}$ is the Fermi distribution function at zero temperature of the third nucleon. Here we assume for $n_{\bm{p}_3}$ a step function approximation. A possible improvement of this treatment, is to consider instead of a step function, a correlated distribution function \cite{baldo+90,li16,arianna13}. 
The total NN interaction $V$ (entering in the Bethe--Goldstone equation (\ref{bg})) is finally given by: 
$V=V_{NN}+V_{eff}/3$. 
The factor $1/3$ is introduced to get the correct normal ordered two-body part of the TNF at Hartree-Fock level  \cite{hebeler10}. 

In order to regularize the density dependent interaction $V_{eff}$, we have used local cutoffs of the form 
\begin{equation}
F_\Lambda=e^{-q^{2n}/\Lambda^{2n}} 
\end{equation}
where $q$ is the exchanged momentum of the two remaining nucleons after the average of the TNF. 
In particular, to regularize the $V_{eff}(\rho)$ associated with the EM potential \cite{N3LO} NN potential we use 
$n=2, \, 3$ for $\Lambda=500, \, 450$ MeV respectively, 
while to regularize the $V_{eff}(\rho)$ associated with the new potential of Ref.\ \cite{maria15,maria_local} 
we use the same form employed for the bare NN interaction, which in momentum space reads: 
$F_{R_S}=e^{-{R_S}^2 q^2 / 4}$ with $R_S=0.7$ fm \cite{maria15,maria_local}.      
We note that for the local cutoffs eployed in this work, a more correct procedure would require to use a three-body regulator directly in Eq.\ (\ref{eq:normord_singpart}); 
this is due to the fact that in this case the regulator is not symmetric under cyclic permutations and 
some additional terms contribute to the average. In the present work we do not consider this aspect 
and we plan to study it in the future.

\section{Results and discussion}

In this section we present and discuss the results of our calculations for the equation of state (EOS), 
i.e. the energy per particle $E/A$ as a function of the density $\rho$, for symmetric nuclear matter (SNM) 
and pure neutron matter (PNM) using the chiral nuclear interaction models and the BHF approach described 
in the previous two sections.  
Making the usual angular average of the Pauli operator and of the energy denominator \cite{gra87,baldo+91}, 
the Bethe--Goldstone equation (\ref{bg}) can be expanded in partial waves. 
In all the calculations performed in this work, we have considered partial wave contributions up to a total 
two-body angular momentum $J_{max} = 8$.  

In Fig.~\ref{fig1} we show the energy per particle of PNM [panel (a)] and SNM [panel (b)] for the 
considered interaction models. The dashed lines, in both panels, refer to the calculations performed employing 
the two-body potential without any TNF, whereas the continuous lines refer to the calculations where 
the contribution of the TNFs to the energy per nucleon has been included.  
Concerning the results for the new local chiral interaction with $\Delta$ isobar degrees of 
freedom \cite{maria_local}, we show in Fig.\ \ref{fig1} the energy per particle relative to the parametrizations  
N3LO$\Delta$+N2LO$\Delta1$ and N3LO$\Delta$ (i.e. without TNF).   
A comparison between the parametrizations N3LO$\Delta$+N2LO$\Delta1$ and N3LO$\Delta$+N2LO$\Delta2$ 
is then shown in a separate figure (Fig.\ \ref{fig1_new}). 

Focusing first on the case of PNM (Fig.~\ref{fig1}(a)), we note sizable differences 
between the three energy per nucleon curves produced by the different NN interactions.   
The model N3LO$\Delta$ (upper (black) dashed line) gives indeed a much stiffer EOS than the N3LO ones 
for both cutoff values, $\Lambda = 500$~MeV (middle (red) dashed line) and 
$\Lambda = 450$~MeV (lower (blue) dashed line).   
This behaviour is both due to the local form of the potential and to the inclusion of $\Delta$ isobar.  

Concerning the role of TNFs in neutron matter, we note that when the original N2LO TNF is reduced to 
an effective density dependent two-body interaction $V_{eff}(\rho)$, using the momentum space average 
proposed in \cite{holt} and used in the present work, the only terms that survive in PNM after the average 
are the ones proportional to $c_1$ and $c_3$.        
Thus calculations using the models N3LO$\Delta$+N2LO$\Delta1$ and N3LO$\Delta$+N2LO$\Delta2$ 
give the same results in PNM because they are not affected by the values of low energy constants 
$c_D$ and $c_E$, and they have the same values for the LECs $c_1$ and $c_3$.  
In addition, looking at Tab. \ref{tab1}, we see that the values of $c_1$ and $c_3$ are very 
similar for the considered models. 
Thus we expect a comparable effect of TNFs on the EOS for PNM.  
This expectation is confirmed by our results. 
In fact, we find that the contribution $\Delta(E/A)_{TNF}$ to the energy per particle of PNM due to 
the inclusion of TNFs at the empirical saturation density $\rho_0=0.16~{\rm fm}^{-3}$ is 
$\Delta(E/A)_{TNF} = 3.49$~MeV in the case of the N2LO$\Delta1$ TNF and 
$\Delta(E/A)_{TNF} = 3.58$~MeV (4.20~MeV) for the N2LO(500) (N2LO(450)) TNF.  
At $\rho = 0.40~{\rm fm}^{-3}$ we find  
$\Delta(E/A)_{TNF} = 29.37$~MeV for the N2LO$\Delta1$ model and  
$\Delta(E/A)_{TNF} = 30.06$~MeV (33.25~MeV) for the N2LO(500) (N2LO(450)) TNF. 

The EOS for symmetric nuclear matter is shown in Fig. \ref{fig1}(b).  
When only two-body interactions are included, models based on the EM N3LO potential \cite{N3LO} 
give unsatisfactory nuclear matter saturation properties.  
More specifically the model N3LO(500) (middle (red) dashed line) gives a saturation point 
($\rho_0 = 0.41$~fm$^{-3}$, $E/A|_0 = -24.25$~MeV),  
whereas the EOS curve for the model N3LO(450) (lower (blue) dashed line) shows no saturation point 
up to density of $\sim 0.5$ fm$^{-3}$. 
The EOS for the N3LO$\Delta$ NN interaction \cite{maria_local} (upper (black) dashed line)
has instead a very different trend. In this case the saturation point turns out to be 
($0.24$ fm$^{-3}$, $-18.27$ MeV).  
Besides the explicit inclusion of the $\Delta$ isobar, this sizable difference 
in the energy per particle of SMN between the new local chiral potential of Ref.\ \cite{maria_local}
and the EM potential \cite{N3LO} is also due to the strong non locality of the EM N3LO potential.  
It is very interesting to note that the behavior found for the N3LO$\Delta$ potential is very similar 
to the one (saturation point $\rho_{0} = 0.23~{\rm fm}^{-3}$, $E/A|_{\rho_0} = -16.43$~MeV) \cite{logoteta15} 
obtained using the Argonne V18 (AV18) interaction \cite{av18}. 

The overall repulsive effect introduced by the inclusion of TNFs produces a significant improvement 
of the calculated SNM saturation point (see the continuous lines in Fig.\ \ref{fig1}(b)) 
with respect to the results described above for the case with no TNFs. 
We find that the contribution $\Delta(E/A)_{TNF}$ to the energy per particle of SNM due to 
the inclusion of TNFs at the empirical saturation density $\rho_0=0.16~{\rm fm}^{-3}$ is 
$\Delta(E/A)_{TNF} = 1.80$~MeV in the case of the N2LO$\Delta1$ TNF and  
$\Delta(E/A)_{TNF} = 6.22$~MeV (6.39~MeV) for the N2LO(500) (N2LO(450)) TNF.  
At $\rho = 0.40~{\rm fm}^{-3}$ we find  
$\Delta(E/A)_{TNF} = 15.69$~MeV for the N2LO$\Delta1$ model and  
$\Delta(E/A)_{TNF} = 45.42$~MeV (49.33~MeV) for the N2LO(500) (N2LO(450)) TNF. 

These results clearly show that in the case of $\Delta$-full chiral nuclear interactions the contribution 
to the energy per particle generated by the TNFs is strongly reduced in comparison to the case  
where the EOS is obtained from $\Delta$-less chiral interactions.  
Our results thus confirm that a $\Delta$-full ChPT for nuclear interactions has an improved convergence 
with respect to a $\Delta$-less ChPT \cite{kaiser98,krebs07}.

In Tab.\ \ref{tab2} we report the calculated values of the saturation points of SNM for the interaction models 
considered in the present work. 
All the models, with the exception of the N3LO+N2LO(500) one, provide reasonable saturation points.  
We want to remark that using non local cutoff to regularize the TNF adopted in conjunction with the two-body 
EM \cite{N3LO} N3LO interaction, the saturation points can be slightly improved. Assuming indeed the same 
cutoff of the bare NN interaction to regularize the effective density dependent NN potential after the average 
of the three-nucleon force, for model N3LO+N2LO(450) the saturation point turns out to be ($0.17$~fm$^{-3}$, $-15.05$~MeV) 
while for model N3LO+N2LO(500) we have found ($0.155$~fm$^{-3}$, $-13.0$~MeV).  
 
In Fig.\ \ref{fig1_new} we compare the energy per particle of SNM obtained using the two parametrizations 
of the N2LO$\Delta$ TNF, namely N2LO$\Delta1$ (black continuous curve) and N2LO$\Delta2$ (purple dashed-dotted curve). 
It is apparent that at low density, up to $\sim0.25$ fm$^{-3}$, the two models produce an almost identical result 
while at a density of $0.4$ fm$^{-3}$ the difference between the energy per particle is of the order of $3$ MeV. Increasing the nucleonic density this difference is expected to increase. This has important implications for the stiffness of the resulting $\beta$-stable equation of state, in particular for astrophysical applications 
such as the study of neutron stars structure.

%
\begin{table*}
\begin{ruledtabular}
\begin{tabular}{cccccc}
Model & $\rho_0$(fm$^{-3}$) & $E/A$ (MeV) & $E_{sym}$ (MeV)  & $L$ (MeV) & $K_\infty$ (MeV) \\
\hline
N3LO$\Delta$+N2LO$\Delta1$     & 0.171  & -15.23   & 35.39    &  76.0   & 190    \\
N3LO$\Delta$+N2LO$\Delta2$     & 0.176  & -15.09   & 36.00    &  79.8   & 176    \\
N3LO+N2LO(500)                 & 0.135  & -12.12   & 25.89    &  38.3   & 153    \\
N3LO+N2LO(450)                 & 0.156  & -14.32   & 29.20    &  39.8   & 205    \\
\end{tabular}
\end{ruledtabular}
\caption{Properties of nuclear matter at saturation density for the various models described in the text. 
In the first column of the table is reported the model name; in the other columns we give the saturation point of symmetric nuclear matter ($\rho_0$), the corresponding value of the energy per particle $E/A$, the symmetry energy, 
its slope $L$ and the incompressibility $K_\infty$. All these values are referred to the calculated saturation density. }
\vspace{1cm}
\label{tab2}
\end{table*}

The energy per nucleon of asymmetric nuclear matter can be accurately reproduced \cite{bl91} 
using the so called parabolic (in the asymmetry parameter $\beta$) approximation 
%
\begin{equation}
      \frac{E}{A}(\rho, \beta) = \frac{E}{A}(\rho, 0) + E_{sym} (\rho) \beta^2 \,.
\label{parab}
\end{equation}
where $E_{sym}(\rho)$ is the nuclear symmetry energy \cite{epja50,baldo16}. 
The nuclear symmetry energy, and particularly its density dependence, is a crucial ingredient 
to determine the proton fraction in $\beta$-stable nuclear matter \cite{bbb97} and ultimately 
it plays an important role to determine the radius and the thermal evolution of neutron stars \cite{steiner06}. 
Using Eq.\,(\ref{parab}), the symmetry energy can be calculated as the difference between the energy 
per particle of pure neutron matter ($\beta = 1$) and symmetric nuclear matter ($\beta = 0$). 

The symmetry energy, calculated within this prescription, is plotted as function of the 
nucleon density $\rho$ in Fig. \ref{fig_esym}. 
In the case of the N3LO+N2LO interaction model, we obtain a symmetry energy which shows a feeble dependence 
on the value of the cutoff $\Lambda$ in all the considered density range. 
For example, at the empirical saturation density $\rho_{0} = 0.16~{\rm fm}^{-3}$, 
we get $E_{sym} = 28.1~(29.5)~{\rm MeV}$ for $\Lambda = 500~(450)~{\rm MeV}$. 
The symmetry energy calculated with the new local chiral potential of Ref.\ \cite{maria_local} 
(continuous and dot-dashed lines in Fig. \ref{fig_esym}) is systematically above and has a larger 
slope with respect to the one calculated with the N3LO+N2LO interaction model. 
In the same figure, we show $E_{sym}$ (triangles) as obtained from recent calculations \cite{dss14} 
of asymmetric neutron-rich matter with two- and three-body interactions determined respectively at 
N3LO and N2LO of the chiral perturbation theory.  
The results of Ref.\ \cite{dss14} have confirmed the validity of the quadratic approximation  
(Eq.(\ref{parab})) for describing the EOS highly asymmetric matter.  
However, it has been recently shown \cite{seif14,kaiser15} that the $\beta^4$ term in the energy per nucleon 
of asymmetric nuclear matter could not be negligible, especially at supranuclear densities, thus having 
sizable influence e.g. on neutron star cooling \cite{steiner06}.
The two bands in Fig.\ \ref{fig_esym} represent the constraints on the symmetry energy obtained 
by Danielewicz and Lee \cite{dan14} using the excitation energies to isobaric analog states (IAS) in nuclei  
(black-dashed band labeled IAS) and with the additional constraints from neutron skin thickness $\Delta r_{np}$ of 
heavy nuclei \cite{roca13,zhang13} (red-dashed band labeled IAS+$\Delta r_{np}$). 
It should be noted that the IAS constraints have been determined up to saturation density $\rho_0$ while at larger density they  have been extrapolated \cite{dan14}. 
The symmetry energy obtained in the present work both for the N3LO$\Delta$+N2LO$\Delta1$ and the 
N3LO$\Delta$+N2LO$\Delta2$ interaction models is in very good agreement with the experimental 
constraints \cite{dan14} reported in Fig.\ \ref{fig_esym}, whereas $E_{sym}$ as calculated with the 
N3LO+N2LO model (with both $\Lambda = 500~{\rm MeV}$ and $450~{\rm MeV}$)  lies slightly below 
the IAS+$\Delta r_{np}$ red-dashed region.


\begin{figure}[t]
\begin{center}
\includegraphics[width=0.75\textwidth]{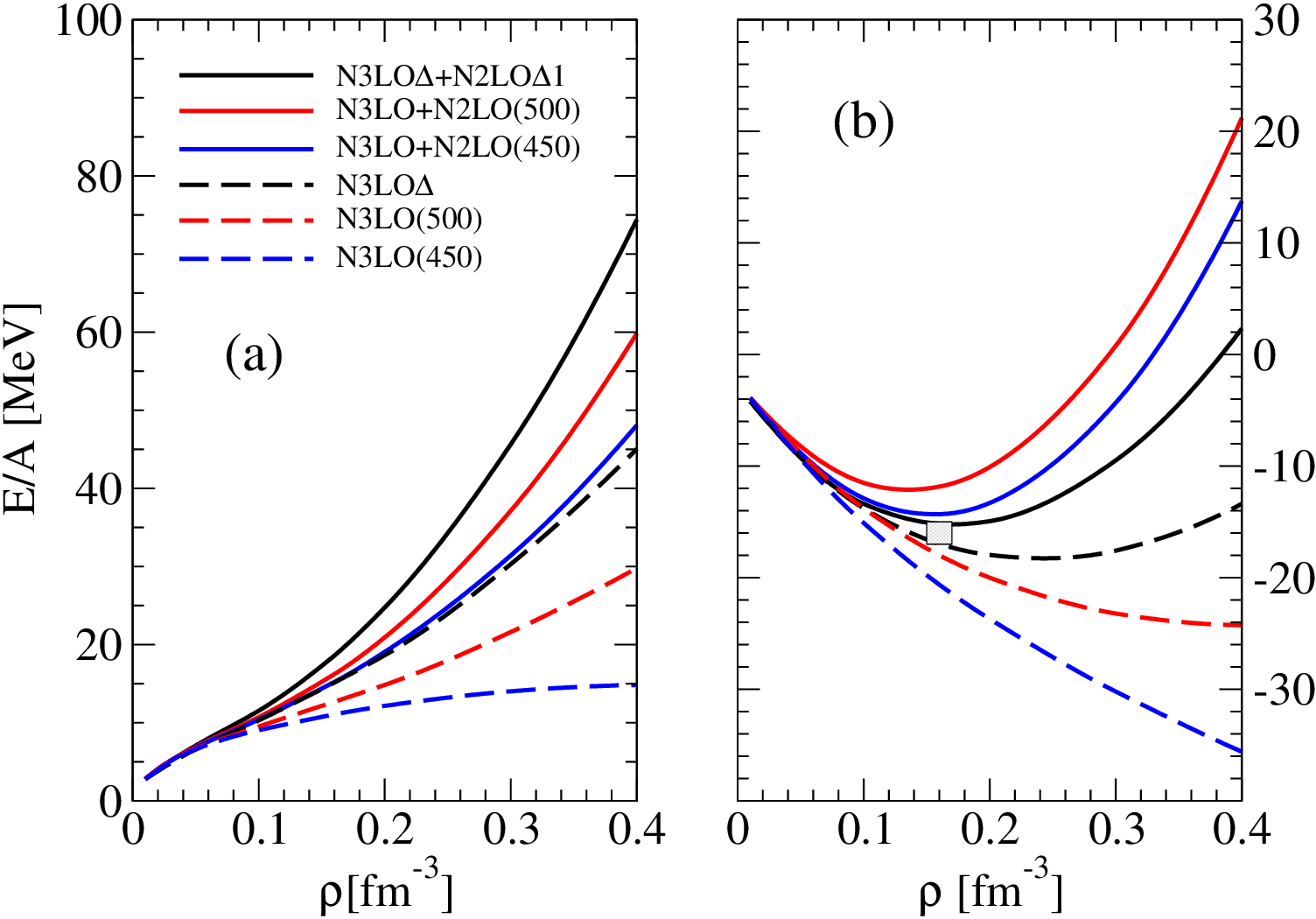}
\caption{(Color online) Energy per particle of pure neutron [panel (a)] and symmetric nuclear matter [panel (b)]  
as function of the nucleonic density for the models described in the text. 
Continuous lines have been obtained using two- plus three-body interactions, 
while the dashed lines have been obtained considering only the two-body interaction. 
The empirical saturation point of nuclear matter  
$\rho_{0} = 0.16 \pm  0.01~{\rm fm}^{-3}$, $E/A|_{\rho_0} = -16.0 \pm 1.0~{\rm MeV}$ 
is denoted by the grey box in the panel (b).} 
\label{fig1}
\vspace{0.5cm}
\end{center}
\end{figure}

\begin{figure}[t]
\begin{center}
\includegraphics[width=0.50\textwidth]{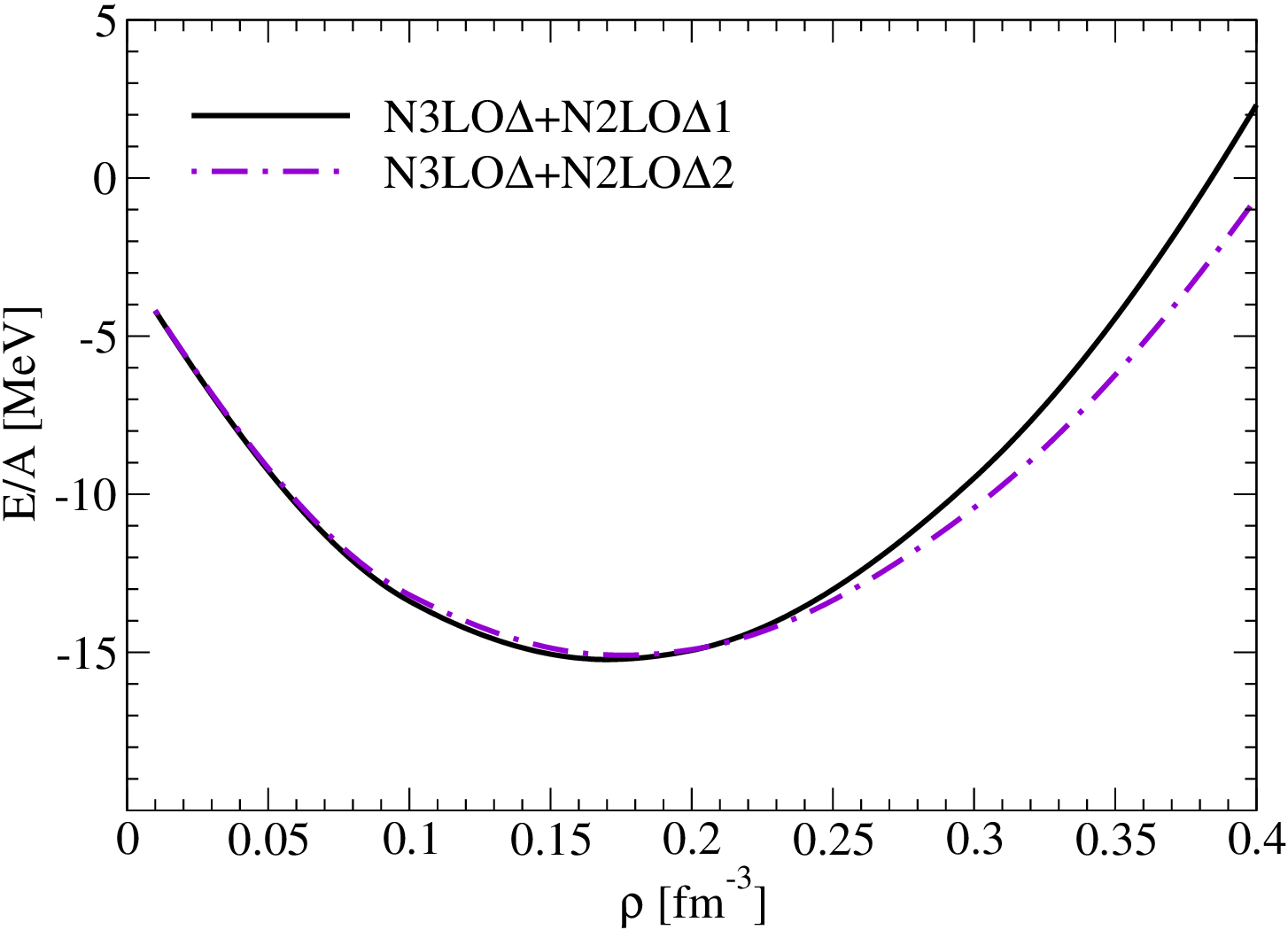}
\caption{(Color on line) Energy per particle of symmetric nuclear matter versus nucleonic density for 
the two TNF models  which take into account the contribution of the $\Delta$ isobar.} 
\label{fig1_new}
\vspace{0.5cm}
\end{center}
\end{figure}

\begin{figure}[t]
\begin{center}
\includegraphics[width=0.50\textwidth]{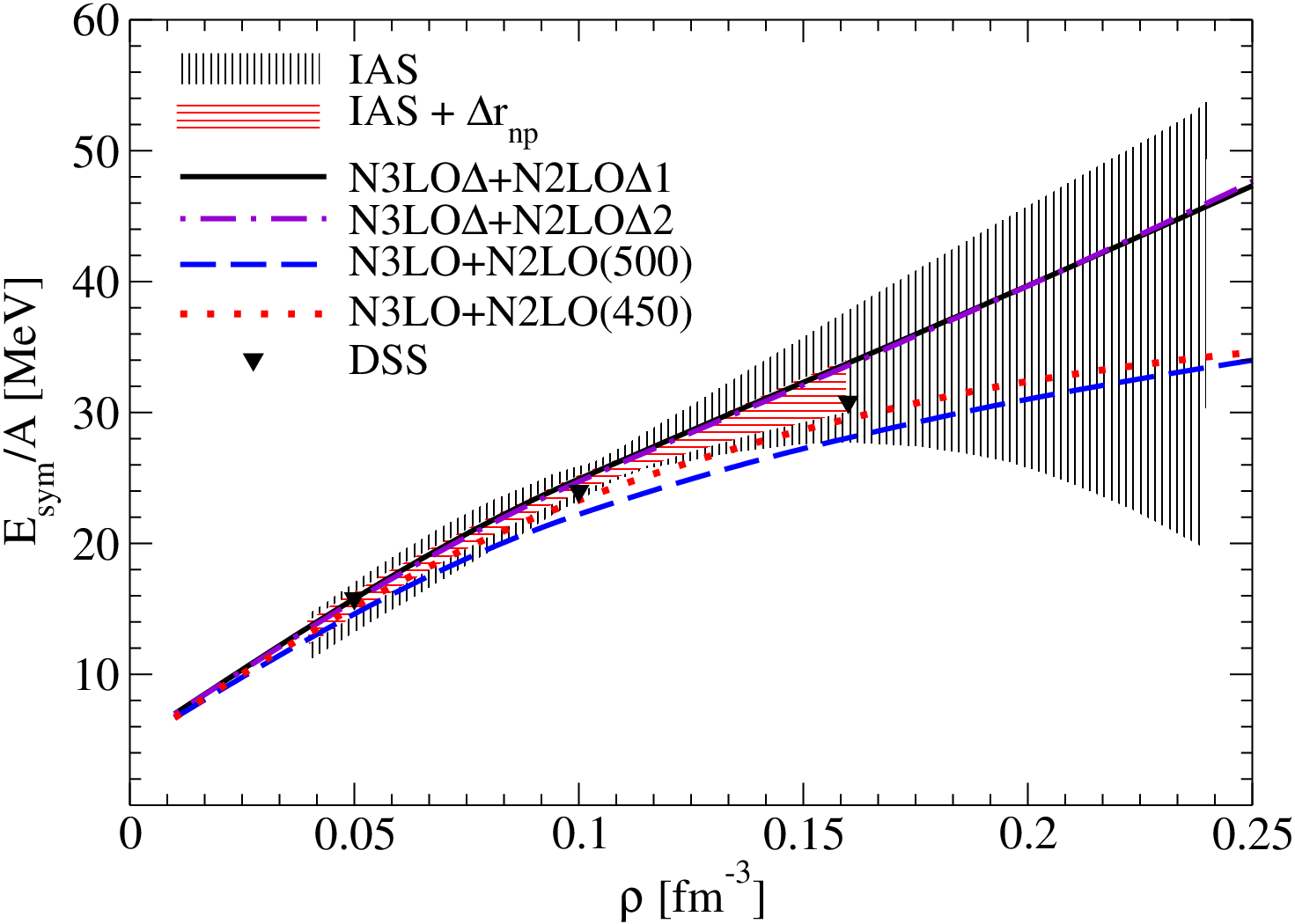}
\caption{(Color on line) Nuclear symmetry energy as a function of the nucleonic density for the four 
interaction models used in the present work.  
The triangles labeled DSS represent the results of Ref.\ \cite{dss14}.  
The black-dashed band, labeled IAS, represents the constraints on the symmetry energy obtained in Ref.\ \cite{dan14} 
using the excitation energies of isobaric analog states (IAS) in nuclei.  
The additional constraints from neutron skin thickness $\Delta r_{np}$ of heavy nuclei give the more limited 
region covered by the red-dashed band labeled IAS+$\Delta r_{np}$ \cite{dss14}. 
}
\label{fig_esym}
\end{center}
\end{figure}

To compare our results with the value of the symmetry energy extracted from various nuclear 
experimental data \cite{epja50,latt14}, we report in Tab.\ \ref{tab2} the symmetry energy 
and the so called slope parameter 
\begin{equation}
       L = 3 \rho_{0} \frac{\partial E_{sym}(\rho)}{\partial \rho}\Big|_{\rho_{0}} 
\label{slope}
\end{equation} 
at the calculated saturation density $\rho_0$ (second column in Tab.\ \ref{tab2}) 
for the interaction models considered in the present work.  
As we can see our calculated $E_{sym}(\rho_0)$ and $L$ are in satisfactory agreement with the values 
obtained by other BHF calculations with two- and three-body interactions 
(see e.g. \cite{ZHLi06,li-schu_08,vidana09,vidana11}) and with the values extracted from various 
experimental data, $E_{sym}(\rho_{0}) = 29.0$ -- $32.7$~MeV, and $L = 40.5$ -- $61.9$~MeV, 
as summarized in Ref.\ \cite{latt14}.       

The incompressibility $K_\infty$ of symmetric nuclear matter at saturation density is given by: 
\begin{equation}
       K_\infty = 9 \rho_{0}^2 \frac{\partial^2 E/A}{\partial \rho^2}\Big|_{\rho_{0}} \,.
\label{incom}
\end{equation}
$K_\infty$ is a very important quantity characterizing the stiffness of the nuclear matter EOS  
with its value having strong implications for the physics of neutron stars \cite{prak97,steiner,Isaac2011} 
and supernova explosions \cite{BKL93,burrows13}. 
The incompressibility $K_\infty$ is usually extracted from experimental data of giant monopole resonance (GMR) 
energies in medium-mass and heavy nuclei. This analysis gives $K_\infty = 210 \pm 30$ MeV \cite{blaizot76} 
or more recently  
$K_\infty = 240 \pm 20$~MeV \cite{shlo06}. 
Recently the authors of Ref.\ \cite{stone10} performed a re-analysis of GMR data in even-even 
$^{112-124}$Sn and $^{106,100-116}$Cd plus earlier data on nuclei with $58<A<208$ finding   
$250$ MeV$<K_\infty<$ $315$ MeV.  
The incompressibility $K_\infty$, at the calculated saturation point for the various interaction models 
used in the present work, is reported in the last column of Tab.\ \ref{tab2}.    
These calculated values for $K_\infty$ are rather low when compared with the empirical values extracted from 
GMR in nuclei. 
This is a common feature with many other BHF nuclear matter calculations with two- and three-body 
interactions (see e.g. \cite{li-schu_08,vidana09}). 

In Fig. \ref{fig4} we compare our results for the energy per particle of PNM [panel (a)] and SNM [panel (b)] 
with those obtained by other researchers using different many-body approaches. 
In the first place in Fig. \ref{fig4}(a) we consider the case of neutron matter.  
The black- and red-dashed regions in Fig. \ref{fig4}(a) represent the results of the many-body perturbative calculations of Ref. \cite{darm13} using complete two-, three- and four-body interactions at the N3LO of the ChPT. 
In particular, the region between the two short-dashed black curves (black-dashed band partially overlapped by the 
red-dashed band) is relative to the N3LO Entem--Machleidt (EM) \cite{N3LO} NN potential, 
whereas the red-dashed band refers to the Epelbaum-Glockle-Mei{\ss}ner (EGM) potentials \cite{EGM}. 
The width of the bands represents the uncertainties related to the values of the LECs and the cutoff of the three- and four-body forces. 
The dot-dashed (magenta) curves in Fig. \ref{fig4}(a), represent the results \cite{AFDMC} 
of an auxiliary field diffusion Monte Carlo (AFDMC) calculation of neutron matter using a local form of 
two- and three-body chiral interactions at N2LO and two different values of the NN cutoff 
(see \cite{AFDMC} for more details). 
The green-dotted line corresponds to the results of Ref.~\cite{wlaz14} with the  
auxiliary field quantum Monte Carlo (AFQMC) obtained with chiral N3LO two-body force plus N2LO TNF. 
As one can see, our results are in very good agreement with all the other calculations considered in 
Fig. \ref{fig4}(a), except with the calculations of Ref.\ \cite{AFDMC} for densities close to the empirical saturation density. 
In fact, in this density region the AFDMC curves are rather "flat" compared to other calculations \cite{AFDMC}, 
thus implying low values, in the range $L = (16.0$--$36.5)$~MeV, for the slope parameter calculated using the 
parabolic approximation Eq.(\ref{parab}) at the empirical saturation density $\rho_0 = 0.16$~fm$^{-3}$.    
We next consider in Fig. \ref{fig4}(b) the case of symmetric nuclear matter. 
The dot-dashed curve labeled N2LO$_{sat}$ corresponds to the energy per nucleon of SNM calculated by the authors 
of Ref.\ \cite{ekstrom} using the Coupled-Cluster method (see \cite{hagen_RPP14} and references therein quoted) 
and the so called N2LO$_{sat}$ interaction \cite{ekstrom}. In this interaction model, two- and three-body chiral 
potentials, at the N2LO of ChPT, have been simultaneously optimized to reproduce low-energy NN scattering data 
as well as the binding energies and radii of few-nucleon systems and of selected carbon and oxygen isotopes. 
Notice that the N2LO$_{sat}$ interaction has the remarkable feature to reasonably reproduce the binding energies and radii of medium-mass nuclei up to $^{40}$Ca  and also to give (see Fig. \ref{fig4}(b)) a satisfactory  
saturation point (0.166~fm$^{-3}$, -14.58~MeV) of SNM.   
The green dashed curve in Fig. \ref{fig4}(b) corresponds to the results of Ref.\ \cite{hebeler11} using 
many-body perturbation theory (MBPT) and adopting the similarity renormalization group (SRG) 
evolution \cite{bogner2010} of the initial NN interaction \cite{N3LO} to produce a soft low-momentum interaction 
so that the convergence of the many-body calculation is greatly accelerated. 
As one can see, there is a satisfactory agreement between our results for the N3LO$\Delta$+N2LO$\Delta1$ interaction and those obtained by the authors of Ref.\ \cite{ekstrom} and \cite{hebeler11}. 

\begin{figure}[t]
\begin{center}
\vspace{1.cm}
\includegraphics[width=0.75\textwidth]{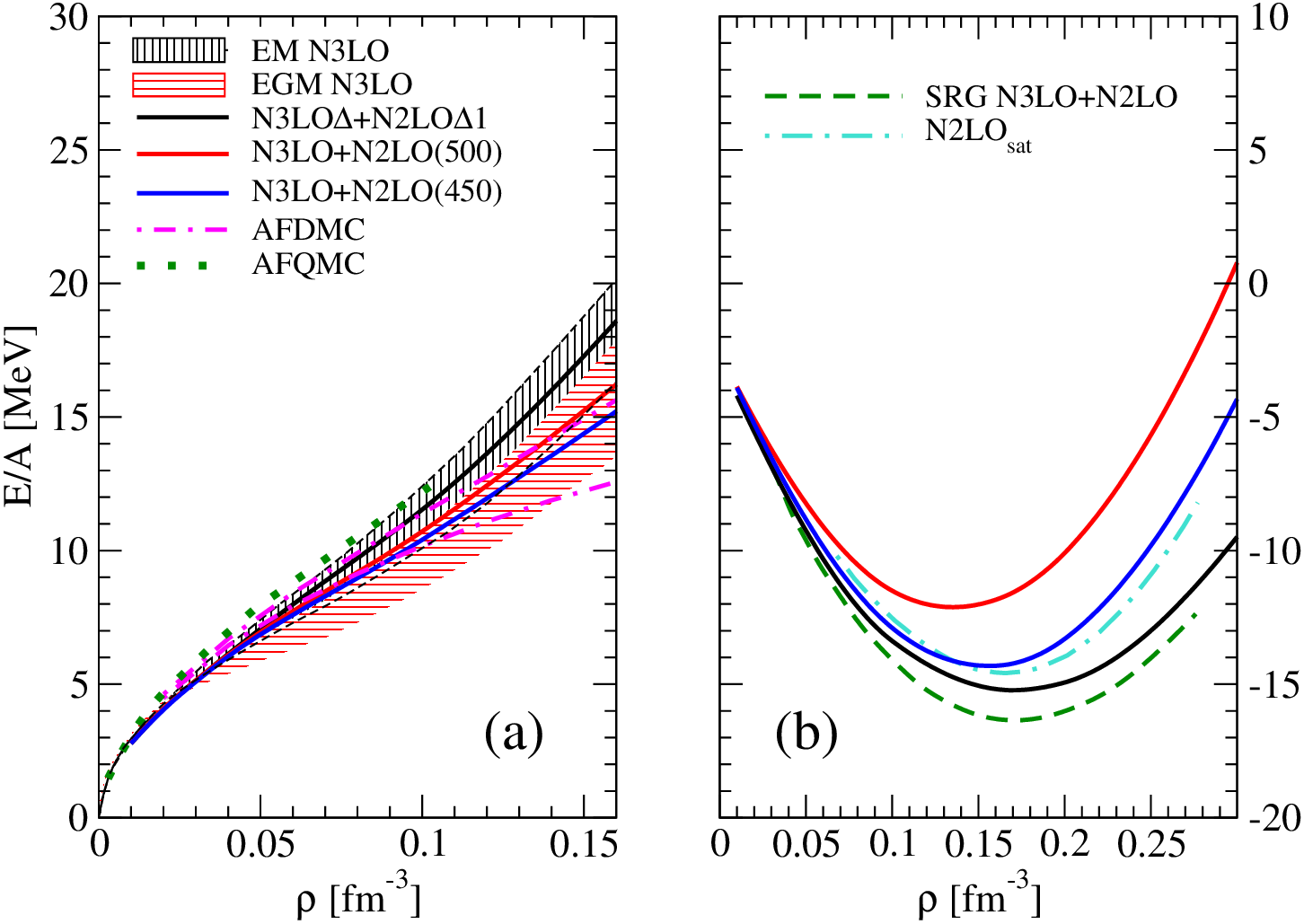}
\caption{(Color online) Comparison of energy per particle of pure neutron [panel (a)] and 
symmetric nuclear matter [panel (b)] between various many-body methods (see text for more details).}
\label{fig4}
\end{center}
\end{figure}

\section{Summary} 

We have investigated the behavior and the properties of nuclear matter using four microscopic models fully based on  interactions derived in chiral effective field theory, in the framework of the Brueckner--Hartree--Fock approach. 
In particular we have tested for the first time the new fully local chiral potential at order N3LO which includes 
the $\Delta$ isobar contributions in the intermediate states of the NN interaction \cite{maria_local}. 
We have also considered two versions of the N3LO chiral NN potential by Entem and Machleidt \cite{N3LO}, 
which differ in the value of the cutoff employed in the calculations.  
All the two-nucleon interactions have been supplemented with TNFs required to produce a good saturation point 
of symmetric nuclear matter.  
For each interaction model we have calculated the energy per particle of SNM and PNM as a function 
of the nucleonic density. From these results we have obtained the saturation point and the incompressibility 
of SNM and in addition the symmetry energy and its slope parameter at the saturation density. 
Our results for these quantities are in good agreement with the available experimental data except for the 
the incompressibility  $K_\infty$ which is underestimated with respect to the highly uncertain empirical 
value \cite{blaizot76,shlo06,stone10}. 
A remarkable agreement has been found with recent experimental predictions for low density behavior of the 
symmetry energy. 
We have found that the inclusion of the $\Delta$ isobar in the NN potential, diminishes the strength of the TNF needed  
the get a good saturation point of symmetric nuclear matter. 
This is consistent with the fact that the chiral effective Lagrangians  
that incorporate the $\Delta$ degrees of freedom allow to a faster convergence of the perturbative series.   
In conclusion the chiral models considered in this work provide solid basis both for the physics light nuclei and low density nuclear matter. 
However a more consistent description would require to go at order N3LO also for what concerns TNFs  
as done in Refs.\ \cite{darm13,drischler16} using MBPT and Self-Consistent-Green's-function (SCGF). 
Such possible extensions will be the focus of future works.  

\section*{Acknowledgments}
We are grateful to Maria Piarulli and Rocco Schiavilla for very usefull discussions.  
This work has been partially supported by ``NewCompstar'', COST Action MP1304.


\end{document}